\documentclass[10pt]{article}

\usepackage[margin=1in]{geometry}
\usepackage{fancyhdr}
\fancypagestyle{plain}{%
\fancyhf{} 
\fancyfoot[R]{-\,\fontsize{8pt}{8pt}\selectfont\thepage\,-} 

}
\usepackage{amsmath}
\usepackage{amsfonts}
\usepackage{amssymb}
\usepackage{graphicx}
\usepackage{sidecap}
\usepackage{xspace} 
\usepackage{ntheorem}
\usepackage{mathtools}
\usepackage{diagbox}
\usepackage{ntheorem}
\usepackage{multirow}
\usepackage{hyperref}

\usepackage{letltxmacro}
\LetLtxMacro{\originaleqref}{\eqref}
\renewcommand{\eqref}{Eq.~\originaleqref}

\newcommand{\dd}{\text d}
\newcommand{\ga}{\alpha}
\newcommand{\gb}{\beta}
\newcommand{\gl}{\lambda}
\newcommand{\equ}[1]{\begin{gather} #1 \end{gather}}

\newcommand{\order}[1]{\mathcal O\left(#1\right)} 
\newcommand{\abs}[1]{\left\vert#1\right\vert} 
\newcommand{\quads}[1]{\quad #1 \quad}
\newcommand{\qand}{\quad \text{and} \quad}
\newcommand{\maxent}{\textsc{m}ax\textsc{e}nt\xspace}
\newcommand{\ipf}{\textsc{ipf}\xspace}
\newcommand{\mle}{\textsc{mle}\xspace}

\newcommand{\config}{\ga}

\newcommand{\set}[1]{\lbrace#1\rbrace}
\newcommand{\prob}[1]{\mathfrak{#1}}
\DeclarePairedDelimiterX\infdivx[2]{(}{)}{
  #1\;\delimsize\|\;#2
}
\newcommand{\infdiv}{D_\textsc{kl}\infdivx}
\DeclarePairedDelimiterX{\condpX}[2]{(}{)}{%
  #1\;\delimsize\vert\;#2%
}
\newcommand{\condp}{\,\condpX}
\theoremstyle{break}
\newtheorem{theorem}{Theorem} 

\newtheorem{lemma}{Lemma} 

\newtheorem{program}{Program}

\newcommand{\multidistro}{\text{mult}}
\newcommand{\dimstatespace}{{\abs{\mathcal{A}}}}
\newcommand{\maxentP}{\hat{\prob p}}

\begin{document}

\thispagestyle{empty}

\begin{flushright}
\phantom{Version: \today} 
\\
\end{flushright}
\vskip .2 cm
\subsection*{}
\begin{center}
{\Large {\bf Categorical Distributions of Maximum Entropy\\[1ex]under Marginal Constraints
} }
\\[0pt]

\bigskip
\bigskip {\large
{\bf Orestis Loukas}\footnote{E-mail: orestis.loukas@uni-marburg.de},\,
{\bf Ho Ryun Chung}\footnote{E-mail: ho.chung@uni-marburg.de}\bigskip}\\[0pt]
\vspace{0.23cm}
{\it Institute for Medical Bioinformatics and Biostatistics\\
Philipps-Universität Marburg\\
Hans-Meerwein-Straße 6, 35032 Germany}

\bigskip
\end{center}

\vspace{5cm}
\subsection*{\centering Abstract}
The estimation of categorical distributions under marginal constraints summarizing some sample from a population in the most-generalizable way is key for many machine-learning and data-driven approaches. We provide a parameter-agnostic theoretical framework that enables this task ensuring (i) that a categorical distribution of Maximum Entropy under marginal constraints always exists and (ii) that it is unique. The procedure of iterative proportional fitting (\ipf) naturally estimates that distribution from any consistent set of marginal constraints directly in the space of probabilities, thus deductively identifying a least-biased characterization of the population. The theoretical framework together with \ipf leads to a holistic workflow that enables modeling any class of categorical distributions solely using the phenomenological information provided.

\newpage 
\setcounter{page}{1}
\setcounter{footnote}{0}
%
%



\section{Introduction}

Most data-scientific problems explicitly or implicitly involve the estimation of a multivariate probability distribution of features from samples of a population. Once determined, such probability distribution allows for computing all interesting measures that describe the population \cite{PhysRevE.102.053314}. For example, classification relies on the conditional probability of a target-feature given the characteristics of predictor-features, which can be readily computed from the joint probability distribution of target- and predictor-features. Another common task in data science is the determination of meaningful pair-wise or higher-order associations to understand the dependency structure among features. For example, it is of interest in epidemiology and related fields to determine the contributions of various  “risk factors” or combinations thereof to the probability of becoming diseased. The presence of associations between risk-factors and a disease can be inferred from generalized odds-ratios, which, in turn, can be readily computed from the joint probability distribution of risk-factors plus disease characteristics.

Current approaches in statistical learning usually start with a parametric model for the joint probability distribution of features that captures the perceived or assumed associations between those features. Parametric modeling has offered convenient ways to approximate and efficiently implement tasks such as classification and feature extraction, thus  becoming state of the art.
In the big-data era, as evidence from larger datasets becomes stronger, more detailed  structures of feature associations become amenable to analysis leading to the design of more complex, less intuitively comprehensible models. This emphasizes two main problems of parametric modeling pertaining to the restriction to probability distributions of the exponential family and the requirement to only choose  parameterizations reflecting independent information obtained about the modeled features.

In this work, we adopt a fully data-driven parameter-agnostic approach to estimate the joint probability distribution of categorical features by casting modeling into an optimization problem in linear algebra. 
The guiding principle of this approach is the maximization of the entropy functional under phenomenological constraints imposed by the provided dataset. 
Starting from the information derived from samples of a population this alternative formulation of modeling deductively leads  to the characterization of the population in the least-biased, most-generalizable way. At an operational level, we employ the old method of iterative proportional fitting \cite{kruithof1937telefoonverkeersrekening,10.1214/aoms/1177731829,10.1214/aoms/1177731604} to estimate probability distributions after incorporating the desired pieces of information from the data by maximizing the entropy functional. 
Naturally formulated in terms of the probabilities of possible configurations of features, \ipf automatically incorporates any relevant information that pertains to pair-wise and, in fact, higher-order associations  without modification of the algorithm, 
augmenting the domain of probability distributions beyond the exponential family. We showcase the combined power of the principle of Maximum Entropy (\maxent) and \ipf by uncovering feature associations in a public dataset from emergency departments in the \textsc{usa} beyond the usually considered pair-wise level.

In summary, working in the space of probabilities not only preserves interpretability at any modeling stage, but considerably enhances the class of distributions from which all desired metrics can be then computed. 
Encoding phenomenological information as a linear system to apply the \maxent principle clearly forms a deductive way to select an optimal solution. At the same time, the flexibility of \ipf helps efficiently tackle the wealth of heterogeneous data encountered in realistic settings.

\section{Problem formulation}

In most data-driven studies, the goal is to estimate a sufficiently generalizable joint probability distribution of the investigated features that should characterize a population using information derived from samples of this population. In our problem formulation and its solution, we concentrate on categorical features making no restriction on the amount while systematizing the form of information that enters the estimation procedure.

Concretely, assume that we are given $L$ categorical features labeled by index $i\in\lbrace1,\ldots,L\rbrace$. Each feature $i$ can take an a priori different number $q_i \in \mathbb N$ of states (evidently $q_i\geq2$) described by state variable $\ga_i\in\{1,...,q_i\}$. 
A particular realization $(\ga_1,\ldots,\ga_L)$ specifying all state variables is called a microstate of the system under investigation. To formalize our notation we introduce the space of microstates
\equ{
\label{eq:StateSpace}
\mathcal A = \left\lbrace (\ga_1,\ldots,\ga_L)~\vert~ \ga_1\in\{1,...,q_1\},\ldots,\ga_L\in\{1,...,q_L\}\right\rbrace
}
of dimension $\dimstatespace=q_1\cdots q_L$. A probability distribution
assigns to each microstate $\vec\ga\in\mathcal A$ some real value $ \prob p(\vec\ga)\in[0,1]$ such that the sum of $\prob p(\vec\ga)$ over $\mathcal A$ remains always normalized to unity.
When clear from  context, we shall use $\ga$ as a vector index to enumerate microstates. In that way, any distribution  $\prob p$ can be regarded as an $\abs{\mathcal A}$-dimensional column vector with entries between 0 and 1, motivating the more general definition of the space of probabilities
\equ{
\label{eq:ProbabilitySpace}
\mathcal P = \bigg\lbrace 
\prob{p}\in\mathbb R^\dimstatespace ~\big\vert~ \prob p_\config \in\mathbb R^+_0 ~\text{ for }~ \ga=1,\ldots,\dimstatespace
\quads{\text{and}} \sum_{\config=1}^\dimstatespace\prob p_\ga=1
\bigg\rbrace
~.
}
Notice that $\prob p\in\mathcal P$ implies $0\leq \prob p_\ga\leq1$ for all $\config=1,\ldots,\dimstatespace$. In the following, any model is automatically understood to be described by a distribution on the probability simplex $\mathcal P$ over the declared mictostates \eqref{eq:StateSpace}. 

Usually, we are given a data matrix $\mathbf X\in\mathcal M_{N\times L}(\mathbb N)$ with the records from a sample of size $N$ in the rows and the features along the columns. Each row of $\mathbf X$ is a microstate vector $\mathbf X_{j,\mathbf\cdot} \in\mathcal A$, whose $i$-th entry gives the state $\ga_i$ in which the $i$-th feature was found in the $j$-th record. Conveniently, the data matrix $\mathbf X$ can be summarized by an empirical distribution $\prob f \in \mathcal P$ of relative frequencies over the microstates,
\equ{
\label{eq:EmpiricalDistribution}
\prob f(\ga_1,\ldots,\ga_L) = \frac{1}{N} \sum_{j=1}^N \delta(\ga_1, X_{j,1}) \cdots \delta(\ga_L, X_{j,L})~,
}
where $\delta(\ga,\gb)$ denotes the Kronecker delta when dealing with categorical variables $\ga,\gb$. 
As we are interested in the associations within sets of features with various cardinalities $\ell\leq L$, the information that enters the estimation of a model distribution $\prob p$ shall be in form of marginal relative frequencies computed from the empirical distribution $\prob f$. In a subset consisting of $\ell$ features labeled by $\set{i_1,\ldots,i_\ell}$ where $1\leq i_1<\ldots<i_\ell \leq L$  with realizations $a \equiv (\ga_{i_1}, \ldots, \ga_{i_\ell})$, a marginal relative frequency 
\equ{
\label{eq:PhenoMoments}
\widehat m_a \equiv \prob f_{i_1,...,i_\ell}(\ga_{i_1},\ldots,\ga_{i_\ell}) =  \sum_{\ga_{i_{\ell+1}}=1}^{q_{i_{\ell+1}}}\cdots \sum_{\ga_{i_L}=1}^{q_{i_L}}
\prob f(\ga_1,\ldots,\ga_L) 
}
is computed by summing the empirical distribution over the remaining $L-\ell$ directions in $\mathcal A$ that do not belong to cluster $\set{i_1,\dots,i_\ell}$. 
Corresponding marginal probabilities $m_a$ can be calculated from any model joint distribution $\prob p\in\mathcal P$ as
\equ{
m_a \equiv \prob p_{i_1,...,i_\ell}(\ga_{i_1},\ldots,\ga_{i_\ell}) =  \sum_{\ga_{i_{\ell+1}}=1}^{q_{i_{\ell+1}}}\cdots \sum_{\ga_{i_L}=1}^{q_{i_L}}
\prob p(\ga_1,\ldots,\ga_L) 
~.
}
The marginal distributions $\prob p_{i_1,...,i_\ell}$ inherit the properties of non-negativity and normalization from $\mathcal P$. Similarly to  the joint distribution, any marginal distribution $\prob p_{i_1,...,i_\ell}$ can be locally represented by a column vector of dimension $q_{i_1}\cdots q_{i_\ell}$ such that all specified marginal probabilities of $\prob p$  can be conveniently encoded as a linear operation
\equ{
\label{eq:MLE:marginalDef}
m_a = \sum_{\ga=1}^\dimstatespace C_{a\ga}\, \prob p_\ga
~.
}
The columns of the binary architecture matrix $\mathbf C\in\mathcal M_{D \times \abs{\mathcal A}}(\mathbb N)$ correspond to the probability directions in $\mathcal P$ indexed by Greek $\ga$ whereas its rows enumerate using Latin index $a$  the $D$ marginal probabilities we wish to incorporate into our model.
Since modeling from empirical data is usually concerned with under-constrained problems where $\text{rank}\,\mathbf C<\dimstatespace$, the matrix $\mathbf C$ is expected to be singular in most applications. In other words, marginal probabilities $m_a$ and design architecture matrix $\mathbf C$ do not unambiguously determine the joint distribution $\prob p$.

In practical applications, only limited information about marginal relative frequencies \eqref{eq:PhenoMoments} in various (possibly intersecting) subsets of features might be accessible, leaving other marginal sums unspecified. Henceforth, we refer to the set of all marginal relative frequencies $\set{\widehat m_a}_{a=1,\ldots,D}$ that are accessible and provided in the problem statement as the summary statistics of empirical distribution $\prob f$.
Naturally, adhering to the empirical summary statistics induces an equivalence class on the simplex $\mathcal P$,
\equ{
\label{eq:EquivalenceClass}
[\prob f] =   \left\lbrace \prob p\in\mathcal P ~\vert~  
\sum_{\ga=1}^\dimstatespace C_{a\ga}\, (\prob p_\ga-\prob f_\ga)= 0
\quad\forall a=1,...,D\right\rbrace
~.
}
By definition, any distribution $\prob p\in\mathcal P$ belonging to the equivalence class shares the same summary statistics with $\prob f$. In the two extreme cases where either the set of marginal probabilities is empty or saturated, it is  $[\prob f]=\mathcal P$ and $[\prob f]=\set{\prob f}$, respectively. Any other summary statistics of $\prob f$ constitutes the realm of non-trivial model building. 

The information contained in the provided set of marginal relative frequencies from empirical distribution $\prob f$
necessarily constrains the space of admissible $\prob p\in\mathcal P$ to the equivalence class $[\prob f]$ inducing a system of coupled linear equations
\equ{
\label{eq:LinearSystem}
m_a\equiv\sum_{\config=1}^\dimstatespace
C_{a\ga}\,
\prob p_\config
\overset{!}{=}
\sum_{\config=1}^\dimstatespace
C_{a\ga}\,
\prob f_\config
\equiv \widehat m_a
\quad\forall\, a=1,...,D~. 
}
At finite $N$, 
the linear problem corresponds to finding all member distributions of the equivalence class $[\prob f]$. Multiplying \eqref{eq:LinearSystem} by $N$ 
gives  a so-called Diophantine system of coupled linear equations. 
More generally when $N\rightarrow\infty$, it becomes a  system of linear equations in $\abs{\mathcal A}$ variables $\prob p_\config\in\mathbb R_0^+$ for each tuple $(\ga_{i_1},..., \ga_{i_\ell})$ in each subset $\set{i_1,\ldots,i_\ell}$. Being an equivalence class in $\mathcal P$ the solution set to linear system  \eqref{eq:LinearSystem} forms a non-empty convex subset of $\mathcal P$, whenever the summary statistics from some empirical distribution $\prob f$ are provided, as it includes at least $\prob f$ itself. 
The binary architecture matrix $\mathbf C$ introduced in  \eqref{eq:MLE:marginalDef} plays the role of the coefficient matrix, while the vector of marginal relative frequencies with components $\widehat m_a$ acts as the  inhomogeneous term. 
The specification of marginal distribution $\prob f_{i_1,\ldots,i_\ell}$ in \eqref{eq:PhenoMoments} in a feature set of cardinality $\ell$ automatically implies  all lower marginal distributions $\prob f_{i_1,\ldots,i_\nu}$ with $\nu<\ell$ ($\nu=0$ being the normalization condition itself). This means that the system of linear constraints described by $\mathbf C$
is generically redundant, whenever the summary statistics of $\prob f$ include marginal distributions on intersecting subsets of features. 

Special care needs to be taken whenever any empirical marginal relative frequency $\widehat m_a$ is exactly zero.
Due to the constraint of non-negativity on the simplex of \eqref{eq:ProbabilitySpace}, $\widehat m_a=0$ automatically sets all probabilities $\prob p_\ga$ contributing to $m_a \overset{!}{=} \widehat m_a$ via \eqref{eq:MLE:marginalDef} to zero:
\equ{
\label{eq:zeroP}
\exists\, a\in [1,D]: ~~
\widehat m_a = 0 \quads{\Rightarrow}
\prob p_\ga = 0 \quad\forall\, \ga\in\{1,\ldots,\dimstatespace\} ~\text{ with }~ C_{a\ga}=1~.
}
We encounter \cite{bishop2007discrete} two types of $a$'s, i.e.\ realizations $(\ga_{i_1}, \ldots, \ga_{i_\ell})$ in a  feature subset $\set{ i_1,\ldots,i_\ell}$, with a marginal relative frequency $\widehat m_a = 0$. One type, referred to as \textit{sampling zero} is due to insufficient sampling of microstates $\ga$ with a relatively small probability participating in the $a$-th marginal sum \eqref{eq:PhenoMoments}. Thus in principle, any sampling zero is expected to disappear by sufficiently increasing the sample size or via regularization. The other type, which is referred to as \textit{structural zero}, is logically known to have zero value, because it corresponds to certain feature realizations that are impossible, e.g.\ being biologically male and pregnant. In contrast to sampling zeros, a structural zero encodes important deterministic relationships between features and their realizations.

As a deterministic constraint, the presence of structural zero(s)  $\widehat m_a = 0$  modifies the architecture of linear system \eqref{eq:LinearSystem}.  
In order to describe the probabilistic constraints applied to the subset $\mathcal A'\subseteq\mathcal A$ which only includes those microstates that are \textit{not} associated with any structural zero, we first define a reduced coefficient matrix $\mathbf C'$ with  all rows corresponding to structural zeros and columns corresponding to the microstates with zero probability removed. From the rank-nullity theorem we know then that the degrees of freedom in $\mathcal A'$ which remain unconstrained by the summary statistics are given by
\equ{
\label{eq:DimProblem1}
\dim\ker \mathbf C' = \vert\mathcal A'\vert - \text{rank}\,\mathbf C'
~,
}
where the rank of the reduced coefficient matrix defines the dimension of the linear problem,  $d \equiv \text{rank}\,\mathbf C'$. As long as the reduced coefficient matrix $\mathbf C'$ implies $d < \vert\mathcal A'\vert$, there exists a non-trivial equivalence class $[\prob f]$ given the summary statistics, so that we anticipate non-trivial model estimation.

\section{The principle of maximum entropy}
\label{sc:MaxEnt}
In the previous section, we have outlined how information about the joint probability distribution of a population in form of summary statistics from its samples
enters the estimation procedure. This yields the linear system in \eqref{eq:LinearSystem} which relates the marginal relative frequencies $\widehat m_a$ to some model $\prob p\in\mathcal P$ via the binary architecture matrix $\mathbf C$. The solutions of \eqref{eq:LinearSystem} span the equivalence class $[\prob f]$ of the empirical joint distribution $\prob f$. Now, we need a mechanism to select one of the distributions $\maxentP\in[\prob f]$ as a ``good'' model for the data-generating process. In more mathematical terms, we need to find a functional of distributions over the  equivalence class $\mathcal[\prob f]$ whose extremization shall yield the desired model distribution $\maxentP$.

Lacking any information about the joint distribution of categorical features other than the number $L$ of categorical features and their different number $q_i$ of states, all we know at this point is the number of microstates $\abs{\mathcal A}$. The uniform distribution $\prob u_\config = \abs{\mathcal A}^{-1}$ is unique in that it encodes only this information and nothing more. Only the uniform distribution guarantees that any permutation in the order of microstates yields the same probability assignment. Any non-uniform probability distribution signifies the presence of additional information that distinguishes between microstates. Next, we are given the hint that we will receive a sample with $N$ records. This information does not give ground to any change in the uniform probability assignment yet. However, we could reason about the probability to receive a sample of size $N$, i.e.\ a data matrix $\mathbf X\in\mathcal M_{N\times L}(\mathbb N)$ summarized by the relative frequency distribution $\prob f \in \mathcal P$ of microstates. As discussed in~\ref{sc:samling}, the probability distribution to receive any such sample from a finite population follows a multivariate hypergeometric or approximately a multinomial distribution $\multidistro(N\prob f;\prob u)$.

A large-$N$ expansion of the multinomial probability distribution yields 
\equ{
\label{eq:Distro_Likelihood_Approx}
\multidistro(N\prob f;\prob u) = 
\frac{N!}{\prod_{\config}(N\prob f_\ga)!} \prod_{\config=1}^\dimstatespace (\prob u_\ga)^{N\prob f_\ga}
=
\exp\left\lbrace N H[\prob f] - N\log\dimstatespace
-\frac{\abs{\mathcal A}-1}{2} \log N
+ \order{1}
\right\rbrace~.
}
The sample-dependent leading term corresponds to Shannon's differential entropy

\equ{
\label{eq:ShannonEntropy}
H[\prob f] = 
- \sum_{\config=1}^\dimstatespace \prob f_\ga \log \prob f_\ga
~,
}
which is a measure of uncertainty about the underlying process that generates data.
Essentially, the entropy functional  measures the opposite of information-theoretic distance  of the uniform distribution $\prob u$
from $\prob f$,
\equ{
H[\prob f] = - \infdiv{\prob f}{\prob u} + \log\abs{\mathcal{A}}
~,
}
where the Kullback–Leibler (\textsc{kl}) divergence or entropy of $\prob f$ relative to $\prob u$,
\equ{
\label{eq:KL_divergence}
\infdiv{\prob f}{\prob u} = \sum_{\ga=1}^\dimstatespace \prob f_\ga\log\frac{\prob f_\ga}{\prob u_\ga}
~,
}
represents a conventional notion of distance between two probability distributions in the world of information geometry~\cite{amari2000methods,Cover2006,ay2017information}.
For any pair of distributions it is non-negative definite on account of Jensen's inequality attaining a global minimum iff the two distributions coincide.
Thus, with information only about the microstate space $\mathcal A$ 
and the sample size $N$ the most likely sample is the one whose empirical distribution $\prob f$ maximizes Shannon's entropy $H[\prob f]$ at the global minimum of the associated \textsc{kl} divergence. Incidentally, the entropy functional also governs the large-$N$ expansion of the multinomial coefficient 
\equ{
\label{eq:GibbsEntropy}
 W[N \prob f] =\frac{N!}{\prod_{\ga=1}^{\dimstatespace} (N\prob f_\ga)! }
= \exp\left\lbrace N H[\prob f]
-\frac{\abs{\mathcal A}-1}{2} \log N
+ \order{1}
\right\rbrace~,
}
which combinatorially describes all possible ways to assign $N$ indistinguishable entities to $\dimstatespace$ attributes reproducing the observed multiplicities $N\prob f$. 

Eventually, we are given a sample of size $N$ with data matrix $\mathbf X\in\mathcal M_{N\times L}(\mathbb N)$ summarized by the empirical distribution $\prob f$ of microstates. As before, we extract information from $\prob f$ in the form of marginal probabilities $\widehat m$ to set up the system of coupled linear equations in \eqref{eq:LinearSystem},  constraining the space of viable distributions $\prob p$ to the equivalence class $[\prob f]$. The probability $p\condp{\prob p}{[\prob f]}$ to obtain a distribution $\prob p \in [\prob f]$ is  accordingly determined at large $N$ by 
\equ{
\label{eq:conditionalLikelihood_LargeN}
p\condp{\prob p}{[\prob f]} = 
\left[\sum_{\prob p'\in[\prob f]}\exp \left(NH[\prob p']\right)\right]^{-1}
\exp \left(NH[\prob p] \right) 
~,
} 
on account of \eqref{eq:Distro_Likelihood_Approx} and \eqref{eq:GibbsEntropy}.
Obviously, the probability $p\condp{\prob p}{[\prob f]}$ is maximal whenever the entropy $H[\prob p]$ is maximized. 
The principle of maximum likelihood (\textsc{mle}) dictates how
to choose among the solutions $\prob p\in[\prob f]$ of the linear system a distribution 
\equ{
\maxentP = \max_{\prob p\in [\prob f]} p\condp{\prob p}{[\prob f]}
\label{eq:MaxEntDistro}
}
that maximizes \eqref{eq:conditionalLikelihood_LargeN}. Hence, the most likely distribution $\maxentP$ in the equivalence class $[\prob f]$ is the one of maximum entropy. Any other distribution in  $[\prob f]$ appears  at sufficiently large $N$ exponentially suppressed. This shows that
\mle coincides at large $N$ and the latest asymptotically with the \textit{principle of maximum entropy} (\maxent;~\cite{jaynes2003probability}), which is also known as the \textit{principle of minimum discrimination information}~\cite{kullback1997information} in the statistics field.

The power of the \maxent principle as a deductive mechanism to unambiguously choose an optimal solution $\maxentP$ in $[\prob f]$ is summarized by 
\begin{theorem}
\label{theorem:MaxEnt}
Among all distributions $\prob p$ in the equivalence class $[\prob f]$ induced by marginal probabilities $m \overset{!}{=} \widehat m$ there always exists a unique probability distribution $\maxentP$ having the maximum entropy in the given equivalence class, $\maxentP = \max\limits_{\prob p\in[\prob f]} H[\prob p]$.
\end{theorem}
\noindent
This theorem, alongside further properties of the \maxent solution are constructively proven in detail in~\ref{ap:Proof}. 
Alternatively to the constructive proof of existence outlined in Appendix, 
Theorem~\ref{theorem:MaxEnt} holds given that $[\prob f]$ is a closed convex subset of $\mathcal P$. $\mathcal P$ is compact, while Shannon's entropy is
a continuous concave function over the simplex.
Hence, $[\prob f]$ is compact and Shannon's
entropy necessarily attains a maximum there. Strict concavity of $H$ ensures the uniqueness.
In addition to the reasoning presented in this paper, one can axiomatically arrive~\cite{shore1980axiomatic,paris1990note,csiszar1991least}  at the \maxent principle as the unique mechanism to perform deduction from a provided dataset in a most minimalistic, least-biased way. 

\section{Iterative proportional fitting}
\label{sc:IPF}
As argued in Section~\ref{sc:MaxEnt}, the \maxent principle selects a ``good'' model distribution $\maxentP$ from the solutions in the equivalence class $[\prob f]$ induced by \eqref{eq:LinearSystem}. Theorem~\ref{theorem:MaxEnt} guarantees that $\maxentP$ exists and is unique giving rise to a deductive mechanism to estimate a model distribution from information in the form of marginal relative frequencies $\widehat m$. In this section, we present the algorithmic procedure of iterative proportional fitting (\ipf) directly applied to the estimation of a model joint distribution $\maxentP$ for categorical features.
This flexible algorithm can estimate model distributions that incorporate any order of associations between features, i.e.\ marginal constraints in subsets of different cardinalities, without any formal modification.
The history of \ipf is long dating back as early as 1937~\cite{kruithof1937telefoonverkeersrekening}. Its breakthrough for estimation of missing cells in contingency tables happened mainly due to~\cite{bishop2007discrete}. In information theory, classical papers such as~\cite{ireland1968contingency,csiszar1975divergence} proved its convergence while exploring~\cite{darroch1972generalized, haberman1974log} possible generalizations.
Extensively, this iterative procedure has been used to estimate~\cite{christensen2006log,rudas2018lectures}  log-linear models from (usually two- and three-dimensional) contingency tables and in the context of matrix balancing in the \textsc{ras} algorithm~\cite{10.2307/2525582}. In addition, it appears~\cite{altschuler2017near} in  Optimal Transport as the Sinkhorn-Knopp algorithm. 

Surprisingly, a large part of the literature in the field of machine learning and statistical modeling remains unaware of this method. Though briefly mentioned in~\cite{phdthesis}, \ipf has not been systematically applied in the context of \maxent modeling or more generally in the inverse Ising/Potts problem~\cite{nguyen2017inverse} and  literature of Boltzmann machines~\cite{tanaka1998mean,loukas2019self},  where Newton-Raphson methods prevail~\cite{jaakkola2000bayesian}. The latter are usually motivated~\cite{malouf2002comparison} within  parametric models. In contrast to the dual parametric formulation (outlined in~\ref{ap:VarForm} for the sake of completeness), \ipf operates exclusively in the space of probabilities $\prob p_\ga$ avoiding any potential singularities in parameter space. 
Merits of exclusively working with the physical degrees of freedom $\prob p_\ga$ are the insensitivity of the iterative algorithm to redundant constraints as well as to the presence of structural zeros \eqref{eq:zeroP}. This enables us to directly consider possibly redundant marginal conditions without the need to pre-determine a minimal set of $d$ (\eqref{eq:DimProblem1}) independent constraints. Such ability is to be contrasted with Newton-based methods where any dependency in the parameters or the presence of structural zeros would lead to a singular Hessian matrix requiring (i) a parametrization that only uses Lagrange multipliers encoding independent constraints and (ii) discarding features involved in structural zeros.

Provided a realistic set of marginal relative frequencies $\widehat m_a$ calculated from the empirical distribution $\prob f$ in \eqref{eq:PhenoMoments} and using the uniform initial estimate $\prob p^{(0)} = \prob u$, the \ipf at the $n$-th cycle updates the vector of probabilities by the iterative application of
\equ{
\label{eq:IPF:MainStep}
\prob p^{(nD+a)}_\ga = 
\prob p^{(nD+a-1)}_\ga \left( \frac{\widehat m_a}{m^{(nD+a-1)}_a} \right)^{C_{a\ga}}
\quad\forall\,\ga=1,\ldots,\dimstatespace
~,
}
where $C_{a\ga}$ are the entries of the binary architecture matrix $\mathbf C$ of linear system \eqref{eq:LinearSystem}. Here, without loss of generality we have arbitrarily fixed some ordering in the provided set $\lbrace \widehat m_a\rbrace_{a=1,...,D}$ of marginal constraints. A choice of ordering could influence the speed of convergence, but not the outcome. 
Model marginals are computed using the estimated components of probability vector from  previous step according to
\equ{
m^{(nD+a-1)}_a = \sum_{\ga=1}^\dimstatespace C_{a\ga}\,\prob p^{(nD+a-1)}_{\ga}
~.
}
The prescription in \eqref{eq:IPF:MainStep} is repeatedly 
applied over all given marginal constraints $\widehat m_a$ until the marginal probabilities  equal the marginal relative frequencies from $\prob f$, i.e.\ until 
\equ{
m_a^{(n^*D+a)} = \widehat m_a \quad\forall~
a= 1,...,D 
}
after $n^*$ cycles, at least within machine precision or some pre-specified tolerance.
Of course, the practical importance of \ipf arises in that it always converges after finitely many steps within the desired accuracy to the \maxent distribution in $[\prob f]$. More rigorously,  we show in~\ref{ap:Proof} the following: 
\begin{theorem}
\label{theorem:IPF}
Starting from
the uniform distribution $\prob u$
the algorithm of
iterative proportional fitting  
always converges 
to the \maxent distribution 
$\maxentP\in [\prob f]$ 
satisfying the provided set of phenomenological constraints from $\prob f$ 
that are iteratively fitted over the procedure.
\end{theorem}
Last but not least, one needs  to acknowledge that \ipf works as an iterative procedure to systematically incorporate constraints with functions more general than categorical distributions going beyond marginal sums.

\subsection{Toy model}

\begin{figure}[t!]
\centering
\includegraphics[scale=1.1,keepaspectratio]{"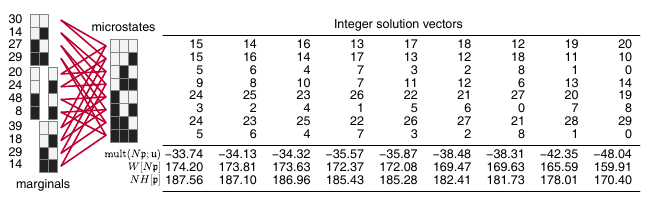"}
\caption{\small From marginal constraints to integer solutions of the linear system: Bipartite graph representation of coefficient matrix $\mathbf{C}$ relating the vector of marginals to the microstate probabilities. On the leftmost, marginal frequencies are provided as input. Boxes refer to feature 1,2, and 3, where white color indicates $\ga_i=0$ and black $\ga_i=1$. Microstates are represented in the same fashion. Table gives the solution set of the Diophantine system consisting of column vectors of counts for all microstates. For each solution vector, three metrics are listed: the logarithm of multinomial probability and multinomial coefficient as well as $N=100$ times entropy.
}\label{fig:Toy:DiophantineSystem}
\end{figure}

To illustrate the application of \ipf to estimate a model distribution $\maxentP$ we start from a simple scenario of $L= 3$ binary features resulting in $\dimstatespace=2^3=8$ microstates $(\ga_1,\ga_2,\ga_3)$ with $\ga_i=0,1$. From a sample of size $N=100$ (of the order often encountered in e.g.\ clinical applications) we obtain a data matrix $\mathbf X$ and its associated empirical relative frequency distribution $\prob f$ and compute marginal relative frequencies $\widehat m_a$ according to \eqref{eq:PhenoMoments}. Choosing some arbitrary  ordering of the marginal constraints $\lbrace \widehat m_a\rbrace_{a=1,...,D}$ the system of coupled linear equations \eqref{eq:LinearSystem} with coefficient matrix $\mathbf C$ incorporating all $D=12$ pairwise marginal constraints takes the form indicated in Figure~\ref{fig:Toy:DiophantineSystem}a. Lines represent 1's in the coefficient matrix $\mathbf C$ dictating which probabilities of microstates participate in the $D$ marginal sums \eqref{eq:MLE:marginalDef}.
Solving the Diophantine system of coupled linear equations \eqref{eq:LinearSystem} we find 9 integer solutions $N\prob p$, which are listed in Figure~\ref{fig:Toy:DiophantineSystem}b. For each of these solutions we calculate the logarithm of multinomial probability $\multidistro(N\prob p;\prob u)$ and of multinomial coefficient $W[N\prob p]$ as well as the entropy $H[\prob p]$ and order them according to the first metric in descending order. 

\begin{SCfigure*}[\sidecaptionrelwidth][!t]
\includegraphics[scale=1,keepaspectratio]{"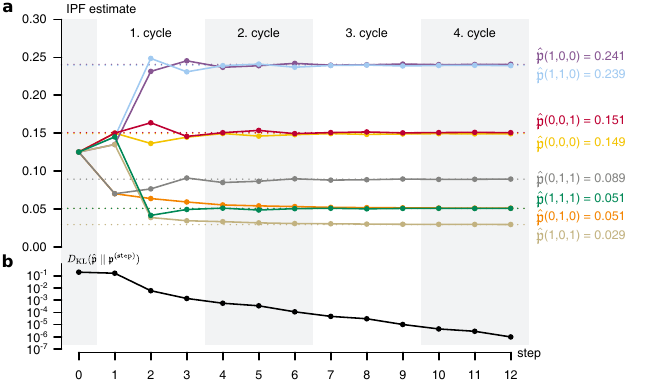"}
\caption{\small Evolution of the estimate for $\maxentP$ during the first 4 cycles of \ipf algorithm:
\textbf{(a)} Probabilities of microstates as estimated by \ipf, starting from the uniform ansatz $\prob u(\ga_1,\ga_2,\ga_3)=1/8$ and fitting empirical relative marginals $\prob f_{ij}(\ga_i,\ga_j)$ deduced from Figure~\ref{fig:Toy:DiophantineSystem} in all pairs of features $\set{1,2}$, $\set{1,3}$ and $\set{2,3}$.  The asymptotic values of $\maxentP(\ga_1,\ga_2,\ga_3)$ are given rounded on the rightmost.
\textbf{(b)} \textsc{kl} divergence of estimated distribution after each step from the limiting \maxent distribution $\maxentP$.}\label{fig:IPF_algorithm}
\end{SCfigure*}

The solution in the top row not only has the highest ($\log$) likelihood among the solution set, but is also the one which can be realized in the most ways displaying the highest entropy. Evidently, the rational-valued $\prob p$ at the top of Table~\ref{fig:Toy:DiophantineSystem}b does not necessarily coincide with the \maxent solution on the simplex \eqref{eq:ProbabilitySpace} under pairwise constraints. Maximizing the entropy functional defined over the one-dimensional solution space associated to \eqref{eq:LinearSystem} in the present setting yields the optimal, real-valued distribution $\maxentP\in[\prob f]$. Its entries are given on the rightmost of Plot~\ref{fig:IPF_algorithm}a. Rounding all entries of the real-valued vector $100\maxentP$ to the nearest integer recovers the most probable solution at the top of Table~\ref{fig:Toy:DiophantineSystem}b. Already at a sample size of $N=100$, the entropy $H[\prob p]$  can well approximate $N^{-1}\log W[N\prob p]$ with a relative error of $\order{10^{-1}}$. 
Hence, our conclusion about $\maxentP$ as the asymptotic solution that is most likely and realized in  the most possible ways remains unaltered. In a generic setting, it quickly becomes~\cite{diaconis1995rectangular} computationally infeasible to determine all solutions of the Diophantine linear system or analytically solve entropy extremization in $[\prob f]$ for $\maxentP$, so that iterative methods directly yielding the limiting distribution become indispensable.

An iterative method based on some parametric ansatz requires to first choose an encoding for the feature realizations such that a non-redundant set of independent parameters can be defined in order to estimate the \maxent distribution $\maxentP$. By contrast, we apply \ipf directly on the $D=12$ marginal relative frequencies in all two-clusters requiring only to arbitrarily fix an ordering of microstates in $\mathcal A$ and of the rational-valued marginal constraints $\lbrace \widehat m_a\rbrace_{a=1,...,D}$. 
Let us first see how to estimate the probability of microstate $(0,0,0)$ for which the  initial uniform probability of $\prob p^{(0)}(0,0,0)=1/8$ deviates from the \maxent value $\maxentP(0,0,0)\approx0.149276$ by more than $16\%$. In the first cycle, we start by applying \eqref{eq:IPF:MainStep} on feature pair $\set{1,2}$:
\small{
\begin{align*}
&\prob p^{(1)}(0,0,0) = \prob p^{(0)}(0,0,0)  \frac{\prob  f_{12}(0,0)}{\prob p^{(0)}(0,0,0)+\prob p^{(0)}(0,0,1)}
= \frac{3}{20}~~,&&~~
\prob p^{(1)}(0,1,0) = \prob p^{(0)}(0,1,0)  \frac{\prob f_{12}(0,1)}{\prob p^{(0)}(0,1,0)+\prob p^{(0)}(0,1,1)}
=\frac{7}{100}~,
\\
&\prob p^{(1)}(1,0,0) = \prob p^{(0)}(1,0,0) \frac{\prob f_{12}(1,0)}{\prob p^{(0)}(1,0,0)+\prob p^{(0)}(1,0,1)}
= \frac{27}{200}
~,&&~~
\prob p^{(1)}(1,1,0) = \prob p^{(0)}(1,1,0) \frac{\prob f_{12}(1,1)}{\prob p^{(0)}(1,1,0)+\prob p^{(0)}(1,1,1)}
= \frac{29}{200}
~.
\end{align*}
}\normalsize
The estimated probabilities of microstates $(0,1,0)$, $(1,0,0)$ and $(1,1,0)$ are needed for the next steps of the cycle.
Proceeding with the subsets $\set{1,3}$ and $\set{2,3}$ we compute
\small{
\begin{align*}
\prob p^{(2)}(0,0,0) =  \prob p^{(1)}(0,0,0) \frac{\prob f_{13}(0,0)}{\prob p^{(1)}(0,0,0)+\prob p^{(1)}(0,1,0)}
= \frac{3}{22}~,
&&
\prob p^{(2)}(1,0,0) = \prob p^{(1)}(1,0,0)  \frac{\prob f_{13}(1,0)}{\prob p^{(1)}(1,0,0)+\prob p^{(1)}(1,1,0)}
= \frac{81}{350}\phantom{\hspace{0.28cm},}
\\
\text{and}&&\prob p^{(3)}(0,0,0) = \prob p^{(2)}(0,0,0)  \frac{\prob f_{23}(0,0)}{\prob p^{(2)}(0,0,0)+\prob p^{(2)}(1,0,0)}
= \frac{273}{1888}
~,
\end{align*}
}\normalsize
respectively. The estimate $\prob p^{(3)}(0,0,0)$ in the last step of the first cycle deviates from the actual \maxent $\maxentP(0,0,0)$ by  $3\%$ improving our naive uniform guess by more than $13\%$.
The \ipf estimation of all $\prob p_\ga$'s over the first four cycles is depicted in Figure~\ref{fig:IPF_algorithm}a. 
As the effect of applying marginal constraints over the first cycle is stronger, quickly becoming small in later cycles, we have used a logarithmic scale to increase visibility.
In the lower part, we plot (also in logarithmic scale) the \textsc{kl} divergence of the \ipf estimate from limiting distribution $\maxentP$ 
to exemplify how the running estimate of the iterative procedure 
approaches the asymptotic distribution. 
After having deduced the \maxent distribution $\maxentP$ from the summary statistics of marginal vector $\widehat m$, one can compute various metrics.
In particular, from the log-odds 
\small{ 
\begin{align*}
&h_1 = \log\frac{\maxentP(1,0,0)}{\maxentP(0,0,0)}\approx 0.478 \quad&& \quad
h_2 = \log\frac{\maxentP(0,1,0)}{\maxentP(0,0,0)}\approx 1.079 \quad&&\quad
h_3 = \log\frac{\maxentP(0,0,1)}{\maxentP(0,0,0)}\approx 0.01 &
\\[1ex]
&J_{12} = \log\frac{\maxentP(1,1,0)}{\maxentP(0,1,0)} - h_1\approx 1.073 \quad&& \quad
J_{13} = \log\frac{\maxentP(1,0,1)}{\maxentP(0,0,1)} - h_1\approx -2.117 \quad&&\quad
J_{23} = \log\frac{\maxentP(0,1,1)}{\maxentP(0,0,1)} - h_2 \approx 0.556&
\end{align*}
}\normalsize
one recovers a familiar-looking description~\cite{croce2019towards} --\,dating back to the well-known Ising model~\cite{ising1925contribution}\,-- of the \maxent distribution defined by pairwise marginal constraints, in terms of interaction matrix $\mathbf J$ and bias vector $\vec h$. Those metrics correspond to the Lagrange multipliers that incorporate a minimal set of $d-1=6$ marginal constraints in the exponential family of distributions.

We have demonstrated that a parameter-free problem formulation together with the compatibly parameter-agnostic \ipf recovers the model distribution $\maxentP$ of maximal entropy that satisfies all marginal constraints provided in the problem statement. Having the optimal solution $\maxentP$ at hand, the optimal values of parameters in \textit{any} meaningful parametrization can be calculated. This clearly demonstrates that the notion of parameters is not required for the definition and estimation of $\maxentP$. Not only are parametric models superfluous, their application is restricted to distributions in the exponential family, rendering the estimation of distributions with structural zeros cumbersome. Thus, our ansatz is much more general than any parametric model. In this context, \ipf becomes a uniquely flexible iterative estimator, because its application only requires a consistent set of marginal constraints in arbitrary subsets of $L$ features, whose information is iteratively incorporated into the model distribution.

\section{Application}
\label{sc:Application}
During the analysis of real-world data we are inevitably confronted with the pitfalls of a finite sample that could distort the picture we deduce about the population the sample descended from. In our setting, a sampled dataset is naturally described by a set of marginal constraints (i.e.\ empirical relative frequencies) from the sample's empirical distribution. Generically, larger sample sizes tend to increase our trust on marginal constraints in larger subsets of features. 
Choosing marginal relative frequencies from the provided data matrix $\mathbf X$ as constraints, automatically selects also candidates for associations, because all unspecified or unimplied marginal constraints lead to the absence of the corresponding associations in the model, i.e.\ the generalized odds ratios correspondingly equal unity. In the following, we shall assume for concreteness that any comprehensive set of marginal probabilities $\widehat m$ calculated from the data $\mathbf X$ consists of  marginal distributions in clusters of features. The cardinality of a cluster defines the order of the marginal distribution which points to the order of association among the features.

In a fully data-driven spirit, there is ground to neither postulate nor exclude any pair-wise or higher-order associations. In principle, one would  need to investigate any possible set of marginal constraints that could be computed from $\mathbf X$. Since the powerset of $L$ features includes $2^L-1$ non-empty clusters, we anticipate at most $2^{2^L-1}$ sets of clusters of features, but symmetries among marginal distributions massively reduce this number. By comparing the \maxent distributions derived from summary statistics over the various sets of constraints, the minimal set of marginal distributions  required to accurately capture all major trends in the population can be found. We demonstrate modeling of real-world data  and propose how to choose a set of constraints at the level of marginal distributions that can lead to the most accurate description of population statistics based on samples of a given size. To select one of the admissible sets of marginal constraints that would lead to a faithful description of the data we use an approach inspired by subsampling~\cite{politis1999subsampling} and the bootstrap~\cite{beran1983bootstrap}.

Specifically, let us consider a fictitious infinite population. Within the state space $\mathcal A$ this ``population'' has (definite and known) microstate distribution coinciding with the empirical distribution $\prob f$ from the provided sample. Thus, we assume an asymptotic distribution, which is in general expected to contain any possible associations. Subsequently, we sample from this ``population'' multiple datasets $s$ 
of various sizes resulting in sample distributions $\prob f^{(s)}$  and compute the \maxent distributions $\maxentP^{(s, k)}$ associated to the equivalence classes $[\prob f^{(s)}]_k$ for all distinct summary statistics of each sample $s$ enumerated by $k$.
Intuitively, our aim is to select the summary statistics $k^*$ which leads to model distribution $\maxentP^{(s,k^*)}$ that is closest to the ``population'' distribution $\prob f$ of microstates. The distance of model distribution  $\maxentP^{(s,k)}$ from $\prob f$ is naturally quantified by the \textsc{kl} divergence 
\equ{
\infdiv{\prob f}{\maxentP^{(s,k)}} = - H[\prob f] - \sum_{\ga=1}^\dimstatespace \prob f_\ga \log \maxentP^{(s,k)}_\ga
~,
\nonumber
}
which essentially measures the opposite of the likelihood of the empirical distribution under the given model (the entropy of the provided dataset being constant). In this context, it becomes reasonable to regularize subsample distributions $\prob f^{(s)}$ by adding a pseudocount of unity to all admissible microstates, as discussed in~\ref{sc:samling}, in order to ensure that the relative frequencies observed in $\prob f$ remain also non-zero in $\prob f^{(s)}$ thus preventing infinite \textsc{kl} divergences. 

To demonstrate the fully data-driven reasoning, we have considered all entries documented over the years 2003\,-\,2018 at the emergency department (\textsc{ed}) by the National Hospital Ambulatory Medical Care Survey
(\textsc{nhamcs}) from the public database of  the \textit{National Center for Health Statistics}. Datasets and documentation can be downloaded from \href{https://www.cdc.gov/nchs/ahcd/datasets_documentation_related.htm\#notices}{https://www.cdc.gov/nchs/} for public use.
We have concentrated on patients whose fate is known, excluding anyone who either left against medical advice or left without being seen or before treatment completion.
In addition, anyone who arrived already dead at the \textsc{ed} was excluded.
In this way, $N=392\,454$ records remained from which $44\,293$ were admitted to hospital. A case is considered critical whenever the patient was admitted to an intensive/critical care unit (\textsc{icu}) or died during the hospital stay resulting in $6\,300$ critical cases out of $44\,293$ hospitalized patients. Here, we encounter a structural zero, because non-hospitalized but critical patients are obviously impossible. 
Despite the fact that structural zeros are often encountered in realistic settings signifying associations that are deterministically forbidden, it is common practice to either adjust a parametric model accordingly to ensure that estimation remains possible or (more alarmingly) features are removed. Our ansatz requires neither a change in the model nor to the data, whenever structural zeros are present.  

In detail, we choose to work with $L = 4+1$ categorical features with $\vec q=\left(6,2,2,2,2\right)$  from the original dataset\footnote{
Following analysis was performed in \textsf{R}~\cite{Rref}. All code can be found at \href{https://github.com/imbbLab/IPF/}{github.com/imbbLab/IPF} alongside a full expose of the methodology.
Of course, the analysis demonstrated in this section can be performed in any other conventional language (e.g.\ Python together with~\cite{2020NumPy-Array,IPFpypi}) used for statistical model building, as well.}: age group, sex, arrival by ambulance and hospitalization augmented by the  dimension of criticality defined in the previous paragraph.
In the microstate space of $\dimstatespace = 6\cdot 2\cdot 2\cdot 2 \cdot 2 = 96$ microstates (the counts are given in Supplementary Table S1), only $\vert\mathcal A'\vert=96-6\cdot2\cdot2\cdot1=72$ are not associated with the aforementioned structural zero. To get a feeling about marginal constraints that reflect actual associations in the population beyond finite sampling effects, we first examine models incorporating the observed marginal distributions in \textit{all} clusters of features of a given cardinality $\ell$, which we refer to as the \textit{order} of summary statistics. 
In principle, we have to estimate 5 \maxent distributions corresponding to summary statistics of orders $\ell = 1,2,3,4,5$, but in the reduced state space $\mathcal A'$ the equivalence class $[\prob f]_{\ell = 4}$ incorporating all fourth-order marginal distributions coincides with $[\prob f]_{\ell = 5}=\set{\prob f}$.
To non-trivially obtain any model distribution $\prob p^{(s,\ell)}$  we apply \ipf using   summary statistics of the desired order. 
\begin{figure}
    \centering
    \includegraphics[scale=0.95]{"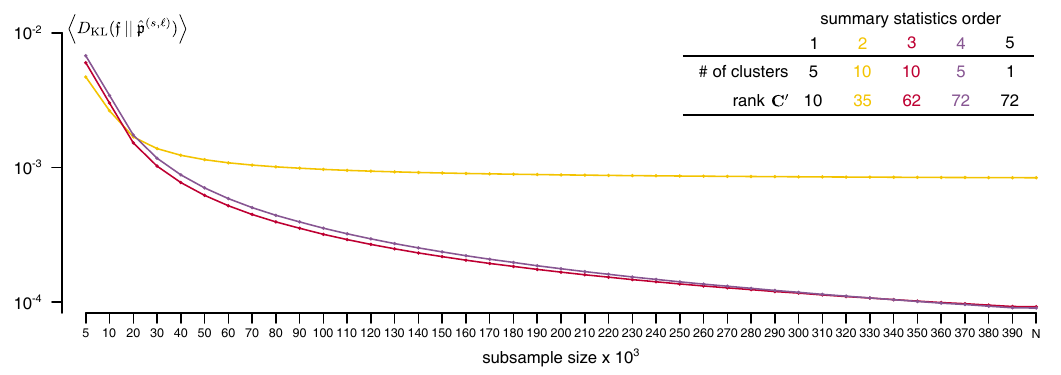"}
    \caption{\small Average \textsc{kl} divergence from empirical distribution $\prob f$ of models  $\maxentP^{(s,\ell)}$ (in $\log$ scale). \maxent distributions $\maxentP^{(s,\ell)}$ are constructed on  $20\,000$ datasets subsampled for each of the sample sizes indicated on the x-axis. The colors indicate models incorporating all two- (yellow), three- (red), and four-order (purple) empirical marginal distributions. The $\ell=1$ model is not displayed for clarity, as it constitutes a poor fit for $\prob f$. The rank of the reduced coefficient matrix is conventionally referred to as the dimension of the model.}\label{fig:uniformOrder}
\end{figure}
After sampling $20\,000$ datasets for each subsample size ranging from $5\,000$ up to $N$, we averaged the \textsc{kl} divergence of the model $\maxentP^{(s,\ell)}$ from ``population'' distribution $\prob f$ over all samples of a given size for the  orders of summary statistics leading to distinct models, as depicted in Figure~\ref{fig:uniformOrder}. 

The average \textsc{kl} divergence for $\ell = 1$ remains at all subsample sizes very large signifying that the corresponding \maxent model cannot accommodate major trends in the empirical distribution as satisfactory as higher-order models. Since summary statistics of order $\ell = 1$ imply independence between the features, this indicates the presence of at least one higher-order association (already anticipated by the presence of structural zero) between the features which is necessary to describe the ``population'' distribution $\prob f$. 
More interestingly, the average \textsc{kl} divergence for $\ell = 2$ at subsample sizes below $20\,000$  was smaller than for $\ell=3,4$. Model distributions $\maxentP^{(s,2)}$ tend on average to describe the ``population'' distribution $\prob f$ better than more complex models at those smaller sample sizes, since there is only limited evidence from the data for higher-order associations. 
Starting at subsample sizes of $20\,000$ up to $340\,000$ the average \textsc{kl} divergences signify that the \maxent distributions deduced from $[\prob f^{(s)}]_{\ell=3}$ constitute optimal fits for the ``population'' distribution $\prob f$. Only at sample sizes larger than $340\,000$ ($>89$\% of the \textsc{nhamcs} original size) the average \textsc{kl} divergence of $\ell = 4$ model becomes the lowest. Notice that even at $\order{N}$ sample sizes the \maxent model deduced from summary statistics of cubic order performs on a comparable scale to the \maxent model deduced from  summary statistics of quartic order, which is the same as directly using $\prob f^{(s)}$. 

In total, we acknowledge that it would have been difficult to identify higher-order associations, although they are clearly present in the ``population'' distribution $\prob f$, if the sample size of the \textsc{nhamcs} dataset were considerably low. On the other hand, at least third-order summary statistics seem to be required given enough data to achieve a better description of the fictitious population, as the pairwise model consistently fails to approach $\prob f$ over a wide range of larger subsample sizes. As the sample size approaches $N$, at least one fourth-order effect may be identifiable. However, the observation that the \maxent model deduced from marginal constraints on all clusters of carnality $\ell=3$ performs almost equally well to the one when $\ell = 4$ suggests that not all quartic-order associations might be needed to faithfully depict the target distribution $\prob f$.

\begin{figure}
\centering
\includegraphics[scale=0.92,keepaspectratio]{"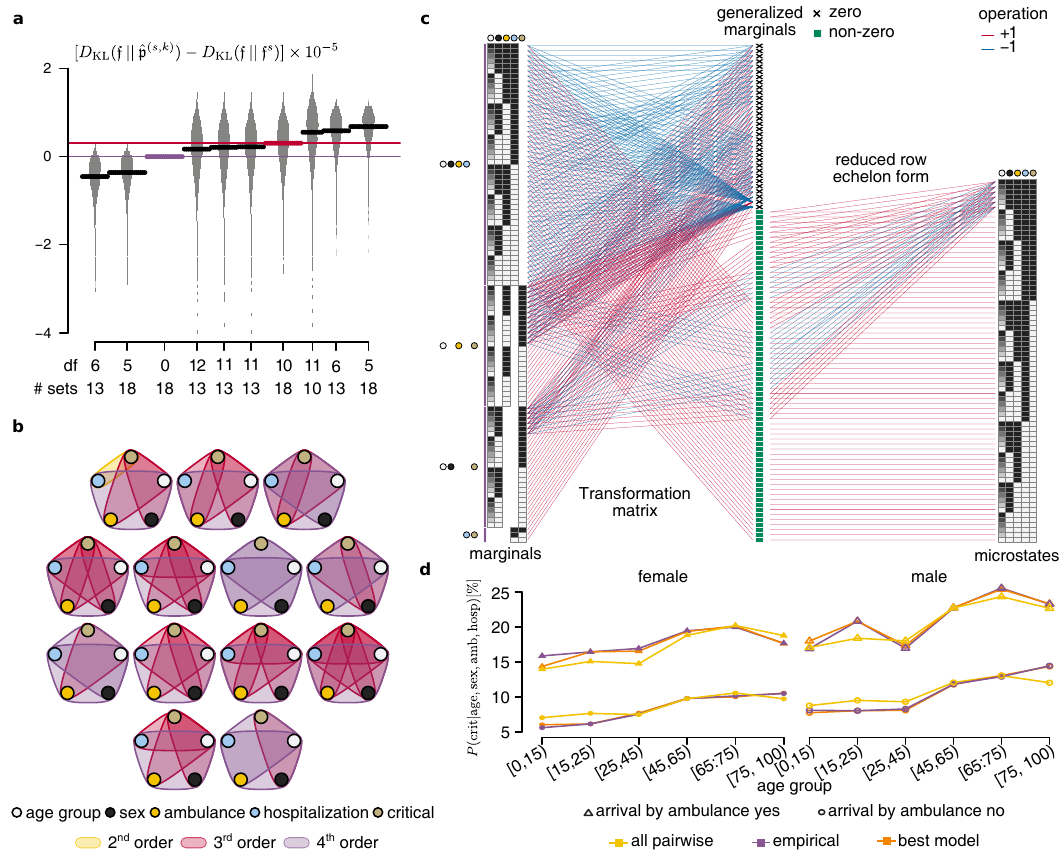"}\caption{Selection of the best fitting model. \textbf{(a)} Difference between the \textsc{kl} divergences from $\prob f$ of a model $\maxentP^{(s,k)}$ and of $\prob f^{(s)}$ is computed over 20\,000 samples. Depicted are the ten top ranking models being closest to $\prob f$. The $\ell=3,4$ models from Figure~\ref{fig:uniformOrder} are signified by red and purple color, respectively.
Each model can be defined via multiple sets of marginal constraints. \textbf{(b)} Hypergraph representation of the set of feature-clusters for which empirical marginal distributions are specified as input. All 13 hypergraphs depicted here lead to identical linear systems and \maxent distribution with the best score in (a). \textbf{(c)} Left column represents the realizations of marginal distributions in the clusters from the first hypergraph in (b). Via Gaussian elimination we obtain the middle column of generalized marginal constraints. The reduced row echelon form relates the non-vanishing generalized marginals (denoted by green) to the probabilities of microstates in the right column. \textbf{(d)} Conditional probabilities given patient's age, sex and means of arrival at the hospital computed from the best-scoring model in (a), the empirical distribution (purple) and the pairwise model (yellow) of Figure~\ref{fig:uniformOrder}.
}\label{fig:mixedOrder}
\end{figure}

To test whether the specification of all empirical marginal distributions in four-clusters is actually required,  we now investigate via our bootstrap scheme all possible summary statistics where each of the $L=5$ features belongs to at least one cluster, resulting in a total of $6\,893$ sets of constraints. 
Calculating each set of marginal distributions $k = 1,\ldots,6\,893$ on each dataset $s=1,\ldots,20\,000$ of size $N$ sampled from the empirical $\prob f$ of the original \textsc{nhamcs} data, we consider again the \textsc{kl} divergence of the deduced \maxent distribution $\maxentP^{(s,k)}$ from $\prob f$. Since the \textsc{kl} divergences of the models in a given sample $s$  are linked via the sample they were estimated from, we compared the \textsc{kl} divergence from $\prob f$ of the \maxent distributions $\maxentP^{(s,k)}$ to the one of the sample $\prob f^{(s)}$ itself, by computing their respective difference. This difference is positive whenever $\prob f^{(s)}$ is closer to $\prob f$ than any non-trivial model $\maxentP^{(s,k)}$ and attains negative values whenever the sample distribution lies further away (Figure~\ref{fig:mixedOrder}a).
Even at sample sizes of order $N$, our analysis reveals that there remain sets of constraints that perform consistently better than trivially using the empirical distribution $\prob f^{(s)}$ of each sample to describe $\prob f$. 

Taking a closer look at the sets of marginal constraints outperforming on average the sample distribution, we detect two groups of summary statistics. Within the first group that scores best for instance, there exist 13 sets of marginal distributions (summarized by hypergraphs in Figure~\ref{fig:mixedOrder}b) resulting in the same \maxent distribution $\maxentP^{(s,k)} = \prob q^{(s)}$ for $k=1,\ldots,13$.
In fact, such degeneracy at the level of hypergraphs representing clusters of features is to be expected and goes beyond the \maxent logic dictating the optimal distribution. The linear systems \eqref{eq:LinearSystem} induced by any of these sets of constraints coincide, as it can be immediately recognized by computing the reduced row echelon form of their coefficient matrices leading to full equivalence of the hypergraphs at the level of the induced equivalence class \eqref{eq:EquivalenceClass}. In principle, the reduced row echelon form of the augmented matrix had to be invoked in order to conclude on the equivalence of linear systems. However, we know that our linear systems generically have by construction infinitely many solutions, hence augmenting the (reduced) coefficient matrix by the vector of empirical marginal sums does not add any new dimension to the span of columns.

Starting from any of the sets of marginal constraints represented by the 13 hypergraphs, it is instructive to distinguish the invariant part defining the same equivalence class that leads to $\prob q^{(s)}\in[\prob f^{(s)}]$ from the redundant specification in terms of marginals (left column  in Figure~\ref{fig:mixedOrder}c depicted for the first hypergraph of \ref{fig:mixedOrder}b). The row reduction of the architecture matrix $\mathbf C'$ that describes one of the sets of redundant constraints correspondingly induces a row reduction of the marginal vector $\widehat m$ leading to a set of $\text{rank}\,\mathbf C' = 66$ generalized marginal constraints (middle column in Figure~\ref{fig:mixedOrder}c) that comprise the linearly independent rules to be obeyed by any $\prob p\in[\prob f^{(s)}]$. Note that in the dual formulation, the non-vanishing generalized marginal constraints would be enforced by Lagrange multipliers (as outlined in~\ref{ap:VarForm}) acquiring the role of parameters.

Having seen that there exist sets of marginal constraints leading to a model distribution $\prob q$ that describes better the ``population'' distribution $\prob f$ starting from its samples $\prob f^{(s)}$ than using $\prob f^{(s)}$ itself, allows us to conclude that any additional empirical constraint at sample sizes of order $N$ cannot be resolved against sampling noise and should be thus omitted.
Once $\prob q\in[\prob f]$ is deduced, it can be used to compute association metrics of biomedical interest, such as conditional probabilities and odds ratios. 
For example, the prediction of critical cases given patient's profile and means of arrival is plotted in Figure~\ref{fig:mixedOrder}d using model $\prob q$ which scores best in the bootstrap. To compare the effect of constraint selection on the predictive power of a model, we also include the same conditional probabilities deduced from the empirical distribution $\prob f$ as well as from the frequently used ``baseline'' model of pairwise summary statistics (yellow model in Figure~\ref{fig:uniformOrder}).
The simpler model of pairwise associations entirely fails to differentiate the age profile by means of arrival (arriving by ambulance results in a mere increase of conditional probability over all age groups by the same amount). Whereas the most complex model induced by $\prob f$ itself seems to mostly perform on a comparable scale to $\prob q$, as anticipated by subsampling scheme in Figure~\ref{fig:mixedOrder}a, it overestimates the risk of arrival by ambulance especially for female younger patients by more than $1.5\%$.

In summary, our problem formulation allowed us to minimally determine sets of phenomenological constraints which go beyond pairwise associations in order to faithfully depict the original dataset. Furthermore, it clearly pointed towards optimal sets of constraints selecting  a model distribution that describes the wealth of associations as good as using the empirical distribution itself, while excluding any effects that are indistinguishable from sampling noise  at the given sample size. 
Analysing the architecture matrix uncovered the degeneracy of hypergraphs in any group of summary statistics which behave identically in our comparison tests, demonstrating that in presence of structural zeros naively enumerating hypergraphs specifying marginal distributions in clusters of features is not enough to filter out all degeneracies. Only linear system \eqref{eq:LinearSystem} gives a conclusive verdict on distinct sets of marginal constraints which are expected to lead to generically distinct \maxent distributions giving a straight-forward classification of inequivalent constraints and hence a robust model definition. 
To arrive at any model distribution neither the data from \textsc{nhamcs} nor the algorithmic procedure of modeling had to be adjusted.
The implementation of \ipf at the level of probabilities $\prob p_\ga$ helps efficiently obtain any distribution given empirical constraints without the need to tackle awkward parametrizations. All in all, the outlined analysis constitutes a purely data-driven approach that can accommodate any class of multivariate categorical distributions with profound consequences in the area of machine learning. \\
\bigskip

\subsection*{Acknowledgments}
{We are thankful to Alan Race for useful discussions and proof-reading the manuscript. Furthermore, we thank Jan-Bernhard Kordaß for useful comments on the original ar\textsc{x}iv version.}
\\\\

\newpage
\appendix

\section{Modeling finite populations}
\label{sc:samling}

\begin{figure}[t]
\centering
\includegraphics[scale=0.4]{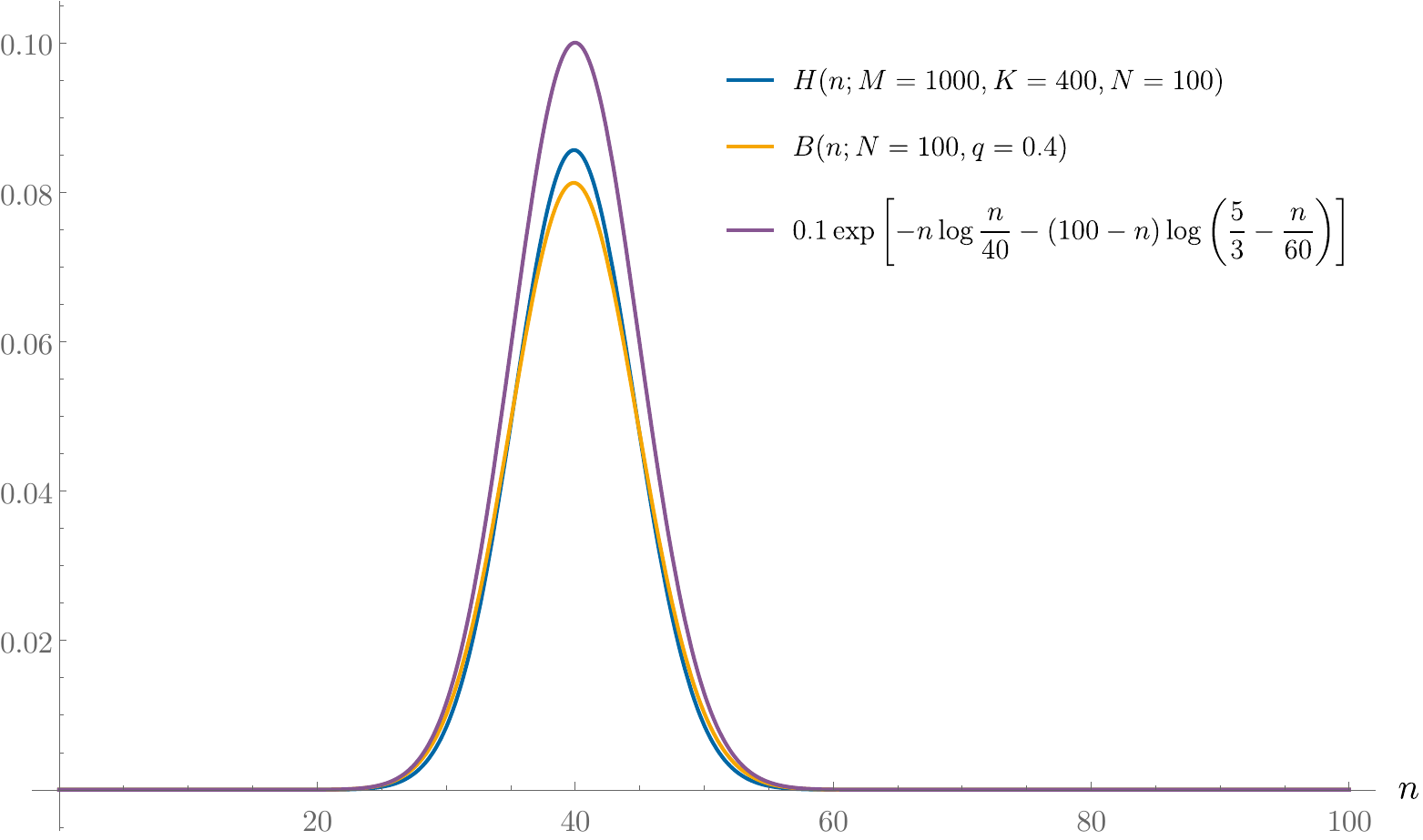}\caption{Comparative plot of the count $n$ of successes as observed in a sample of size $100$ following a hypergeometric distribution $H$  as well as its binomial counterpart $B$ with  the information-theoretic approximation \eqref{eq:MultinomialDistro_Approx}. In the population of $1\,000$ individuals there are $400$ cases of success.}\label{plt:DistrosCompare}
\end{figure}

In the setting of categorical features, observed counts $N \prob f_\ga$ are to be sampled under some model $\prob q$ via a hypergeometric distribution. 
For sufficiently large populations, it is permissible~\cite{hajek1960limiting} to use instead of the hypergeometric  the multinomial distribution.
Treating a population as a pool with infinitely many individuals to sample from is evidently
an idealization. For the purposes of modeling and benchmarking however, this simplifying approximation is mostly sufficient. 
To convince ourselves  we consider in Plot~\ref{plt:DistrosCompare} the simplistic example of a two-dimensional space $\mathcal A=\lbrace\text{success},\text{failure}\rbrace$ with $K=400$ cases of success in a finite population of $M=1\,000$ individuals. We perform an experiment with $N=100$ trials and ask about the observed count $n$ of successes. Immediately, we  recognize that hypergeometric distribution $H(n;M,K,N)$ when sampling \textit{without} replacement and binomial  distribution $B(n;N,q)$ when sampling \textit{with} replacement using $q=K/M=2/5$ exhibit very similar profiles, even in a  smaller population. To compare with the entropy-based analysis advocated in this paper, we also plot (in purple) the information-theoretic approximation to multinomial distribution,

\equ{
\label{eq:MultinomialDistro_Approx}
\log\multidistro (N \prob f; \prob q) =
- N \infdiv{\prob f}{\prob q}
-\frac{\abs{\mathcal A} - 1}{2} \log N
+ \order{1}
~.
}
in terms of the \textsc{kl} divergence of model $\prob q = (q, 1-q)$ from the observed counts $N\prob f = (n, N-n)$.

\paragraph{Pseudo-count regularization.}
In real life, data usually comes at a limited number of samples with the result that certain microstates in $\mathcal A$ with low probabilities remain unobserved in a finite dataset of size $N$, especially whenever the distribution  strongly picks around some other region in microstate space. 
To avoid erroneous singularities in information-theoretic metrics such as the \textsc{kl} divergence, it thus becomes eminent the need to regularize so that the data-driven analysis captures --\,as much as possible\,-- the physically relevant behavior of the underlying system. 
Theoretically, it is reasonable to assume that any configuration in $\mathcal A'$ is possible, $\prob p_\ga\neq 0$ if not otherwise stated, anticipating finite associations among features. To roughly compensate for unobserved  configurations, so-called \textit{sampling} zeros, one can employ~\cite{MorcosE1293} the pseudo-count method. Following the logic of a least-biased setup each configuration $\ga$ (except for those associated to any \textit{structural} zero) uniformly receives a \textit{pseudo}-count, before seeing the actual data. In other words, one works with a slightly  modified empirical distribution
\equ{
\prob f_\config \quads{\rightarrow} \frac{1}{1+\gl/N\cdot \abs{\mathcal A'}} \left[\gl / N + \prob f_\config\right]
}
for some real $\gl>0$, which in turn implies for its summary statistics 
\equ{
\widehat m_a  \quads{\rightarrow}
\frac{1}{1+\gl/N\cdot \abs{\mathcal A'}} 
\left[\frac\gl N \sum_{\ga=1}^{\abs{\mathcal A'}} C'_{a\ga} +  \widehat m_a\right]
~.
}
The effective regularization strength $\gl/N$ is a --\,hopefully small\,-- hyper-parameter
to be tuned depending on the application field.
Generically for larger data sets of size $N$,  it can be taken at order $1/N$.
This simple prescription helps us avoid any erroneous (near)-singular behavior that would appear whenever some entry of marginal vector $m_a$ happens to be close to zero -- meaning that all affected relative frequencies $\prob f_\ga$ and estimates of \ipf update rule  thereof, would be immediately set to zero.

\paragraph{Models with uniform summary statistics.}
Let us give an explicit formula for the dimension of the linear problem $d=\text{rank}\,\mathbf C$ in absence of any structural zero marginal relative frequency $\widehat m_a$, i.e.\ when $\mathcal A'=\mathcal A$.
For concreteness, we uniformly specify marginal distributions in all
\equ{
\sum_{a=1}^D C_{a\ga} = \begin{pmatrix}
L\\
\ell
\end{pmatrix}
\nonumber
} 
subsets of cardinality $\ell$ leading to 
\equ{
D = 
\sum_{i_1=1}^L q_{i_1} \sum_{i_2>i_1}^L q_{i_2}  \cdots \sum_{i_\ell>i_{\ell-1}}^L q_{i_\ell}
\nonumber
}
--\,in principle redundant\,-- constraint equations.
Considering pairwise summary statistics for example, the set of induced marginal constraints in a two-cluster $\{i,j\}$  explicitly constitutes $q_i\times q_j$ redundant equations. Summing either over the $j$-th or $i$-th direction leads to the constraint equations corresponding to one-feature marginal distribution $\prob f_{i}(\ga_i)$ and $\prob f_{j}(\ga_j)$, respectively. However, these one-site constraints would be also obtained from summing  over any other two-feature marginal distributions $\prob f_{ik}(\ga_i,\ga_k)$ and $\prob f_{jk}(\ga_j,\ga_k)$ with $k\neq i, j$.
To avoid overcounting constraints we thus need to effectively reduce the Potts order $q_i$ of features. 
Indeed, regarding one-feature marginal distributions as given for $i,j=1,\ldots, L$ only $(q_i-1)\times(q_j-1)$ from the $q_i\times q_j$ constraint equations  in the cluster $\set{i,j}$ are independent. In turn,  there are only $q_i-1$ independent constraints induced by $\prob f_i(\ga_i)$, as one is implied by the others in conjunction with normalization on the simplex $\mathcal P$.
Counting independent constraints in that way readily generalizes to $\nu$-th order summary statistics, where 
$\left(q_{i_1}-1\right)\times\cdots\times\left(q_{i_\nu}-1\right)$ independent conditions in each subset $\lbrace i_1,...,i_\nu \rbrace\subset\set{1,...,L}$ are anticipated. 
In total, out of the $D$ consistent equations only 
\equ{ 
\label{eq:DimProblem2}
d = 1 + \sum_{\nu=1}^\ell 
\sum_{i_1<...<i_\nu} 
\left(q_{i_1}-1\right)\cdots \left(q_{i_\nu}-1\right) 
\nonumber
}
constraints are independent representing independent information from empirically observed data matrix $\mathbf X$.

\section{Convergence of IPF and unique existence of MaxEnt distribution}
\label{ap:Proof}

In the first section of this paper, we have seen that the phenomenological constraints imposed by marginal relative frequencies from a data matrix $\mathbf X$ together with the optimality goal of \mle defined a linear program.
Summarizing:
\begin{program}
\label{problem:MaxEnt_LP}
The \maxent distribution $\maxentP$ is obtained as the optimal solution to linear program
\equ{
\min_{\prob p\in\mathbb{R}^{\abs{\mathcal A}}} \sum_{\ga=1}^\dimstatespace \prob p_\config \log \prob p_\config
\nonumber
\quads{\text{subject to}}
\sum_{\ga=1}^\dimstatespace C_{a\ga}\, \prob p_\config = \widehat m_a ~\text{ for }~ a=1,...,D
\qand
\prob p_\ga \geq 0  ~\text{ for }~\config=1,...,\dimstatespace~.
\nonumber
}
\end{program}
Evidently, the feasible region of linear program~\ref{problem:MaxEnt_LP}, coinciding with the equivalence class $[\prob f]$ of all distributions compatible with marginals $\widehat m$,  is non-empty provided a consistent set of phenomenological constraints. 
Next, we  turn to optimality to assert the asymptotic existence of the \maxent solution.
Note that uniqueness immediately follows from existence, as entropy is strictly concave (equivalently $\infdiv{\prob p}{\prob u}$ is strictly convex) in $\prob p \in[\prob f]$ and hence possesses at most one global maximum (equivalently minimum of \textsc{kl} divergence) in the convex set spanned by the equivalence class. All in all, we have
\begin{theorem}
Among all distributions $\prob p$ in the equivalence class $[\prob f]$ induced by marginal probabilities $m \overset{!}{=} \widehat m$ there always exists a unique probability distribution $\maxentP$ having the maximum entropy in the given equivalence class, $\maxentP = \max\limits_{\prob p\in[\prob f]} H[\prob p]$.
\end{theorem}
Showing the existence of \maxent distribution in $[\prob f]$ amounts to proving the convergence of \ipf algorithm to $\maxentP$. 
In order to do so we need a base Lemma from probability theory which 
says  that
\begin{lemma}
\label{lemma:PythagoreanIdentity}
\text{If $\maxentP\in\mathcal P$ is \maxent distribution in an equivalence class $[\prob f]$, then}
\begin{equation*}
H[\maxentP] \geq H[\prob p] ~\forall\, \prob p\in[\prob f]
\quads{\Leftrightarrow}  
\infdiv{\prob p}{\maxentP} = H[\maxentP] - H[\prob p]
\quads{\Leftrightarrow}
\sum_{\ga=1}^\dimstatespace \left(\prob p_\config-\prob \maxentP_\config\right) \log \maxentP_\config = 0~.
\end{equation*}
\end{lemma}
In fact, this is a special case of the analogue of 
Pythagorean theorem for \textsc{kl} divergences, see e.g.~\cite{csiszar2004information,nielsen2018information}. 
To prove the forward direction of the equivalence we introduce a family of parametric functions on $\mathbb R$ via
\equ{
\prob s_\config(\gl) = \gl \prob p_\config + \left(1-\gl\right) \maxentP_\config
\quads{\text{for}} \ga=1,...,\dimstatespace~.
}
For each microstate from $\mathcal A$ our function $\prob s_\config(\gl)$ describes a line with slope $\prob p_\config-\maxentP_\config$ and either positive   or zero intercept $\maxentP_\config$. 
In the latter case, $\ga$ belongs to the complement of realization space $\mathcal A'$ and 
the slope becomes also zero, hence $\prob s_\config(\gl)=0$ for any $\gl$.
In the former scenario of positive intercept, we know by continuity of the straight line  that there is always a finite region around $\gl=0$ where $\prob s_\ga$ remains non-negative.
In total, there should always exist some $t>0$ such that 
$\prob s_\ga(\gl)$ is non-negative when $\gl\in(-t,t)$ for all realizations in $\mathcal A$ indexed by $\ga$. Furthermore, being a linear combination of $\maxentP,\prob p\in[\prob f]$ the  vector $\prob s(\gl)$ parametrized by $\gl\in(-t,t)$  automatically satisfies the linear constraints and represents thus a distribution $\prob s(\gl)\in[\prob f]$.
From Lemma~\ref{lemma:PythagoreanIdentity} we then have 
\equ{
H[\maxentP] \geq H[\prob s(\gl)]
}
for those $\gl$ around $\gl=0$ where $\prob s_\ga(\gl)\geq0~\forall\ga$, 
implying in particular that 
\equ{
f(\gl) \equiv \sum_{\ga=1}^\dimstatespace \prob s_\config(\gl)\log \prob s_\config(\gl) - \sum_{\ga=1}^\dimstatespace \maxentP_\config\log \maxentP_\config \geq 0
~.
}
The strictly convex function $f$ attains its minimum at $\gl=0$ 
as $\maxentP$ is assumed to be \maxent in $[\prob f]$. By the derivative condition we immediately get the desired identity:
\equ{
0 \overset{!}{=} \left.\frac{\dd f(\gl)}{\dd\gl}\right\vert_{\gl=0} = 
\sum_{\ga=1}^\dimstatespace \left(\prob p_\config-\maxentP_\config\right) \log \maxentP_\config
~.
}
Proving the opposite direction is more straight-forward. Starting from the re-arranged expression
%
\equ{
-\sum_{\ga=1}^\dimstatespace \maxentP_\config\log \maxentP_\config = - \sum_{\ga=1}^\dimstatespace \prob p_\config\log \maxentP_\config
}
we subtract from both sides $H[\prob p]$ so that 
\equ{
H[\maxentP] - H[\prob p] = \infdiv{\prob p}{\maxentP}\geq 0
}
with equality only when $\prob p=\maxentP$. Thus, $\maxentP$ has the maximum entropy in $[\prob f]$.

In this work, we have advocated the use of \ipf algorithm to obtain the \maxent distribution from empirical marginal constraints working exclusively in the space of physically meaningful probabilities. 
At the $(n+1)$-th cycle after fitting onto marginal distribution in the subset $\set{i_1,\ldots,i_\ell}$ the \textsc{kl} divergence of  \ipf estimate
from empirical distribution $\prob f$ can be written in the closed form
\equ{
\infdiv{\prob f}{\prob p^{(nD+q_{i_1}\cdots q_{i_\ell})}} = \infdiv{\prob f}{\prob p^{(nD)}}  - 
\infdiv{\prob f_{i_1,\ldots,i_\ell}}{\prob p^{(nD)}_{i_1,\ldots,i_\ell}} 
~,
}
where the latter term is also a \textsc{kl} divergence in the space of marginal distributions and thus by Gibbs inequality non-negative. 
This means that the \textsc{kl} divergence of the \ipf estimate from $\prob f$ decreases after fitting onto the marginals in each subset of features.
Since $D_\textsc{kl}$  is bounded from below, the sequence of \textsc{kl} divergences  induced by \ipf given the phenomenological constraints has to attain  its infimum associated to the \maxent solution, as we verify below, the latest at $n\rightarrow\infty$. Convergence is the key power of this algorithm: 
\begin{theorem}
\label{theorem:IPF}
Starting from the uniform distribution $\prob u$ the algorithm of iterative proportional fitting  always converges to the \maxent distribution 
$\maxentP\in [\prob f]$ satisfying the provided set of phenomenological constraints that are iteratively fitted over the procedure.
\end{theorem}
Now, we rigorously prove\footnote{There are various~\cite{10.2307/2528683,10.2307/2242759} ways to prove the convergence of the algorithm, even geometrically~\cite{doi:10.1080/01621459.1970.10481117}. For a detailed  proof on the existence of the optimal distribution and convergence of \ipf to the latter in more general settings see~\cite{csiszar1975divergence}.} Theorem~\ref{theorem:IPF} ensuring automatically the existence of \maxent distribution in the equivalence class, hence completing the proof of Theorem~\ref{theorem:MaxEnt} as well. 

First, we need to assert that the algorithm always converges to a distribution within the equivalence class. Notice  that for any probability distribution $\prob p$  satisfying the given set of linear constraints, 
\equ{
\sum_{\ga=1}^\dimstatespace C_{a\ga} \prob p_\ga = m_a \overset{!}{=}\widehat m_a
\nonumber
}
we have the identity 
%
\equ{
\label{eq:IPF:VanishingRelation}
\sum_{\config=1}^\dimstatespace \log\frac{\prob p^{(nD+a)\phantom{-1}}_\config}{\prob p^{(nD+a-1)}_\config} \left[\prob p_\config - \prob p^{(nD+a)}_\config\right] = 
\left(\log\widehat m_a - \log m^{(nD+a-1)}_a\right)
\sum_{\config=1}^\dimstatespace C_{a\ga}\left[\prob p_\config - \prob p^{(nD+a)}_\config\right]
= 0
~.
}
For the ratio of two consecutive \ipf estimates we have substituted  update rule 
\equ{
\label{eq:IPF_update}
\prob p^{(nD+a)}_\ga = 
\prob p^{(nD+a-1)}_\ga \left( \frac{\widehat m_a}{m^{(nD+a-1)}_a} \right)^{C_{a\ga}}
\quad\forall\,\ga=1,...,\dimstatespace
~.
}
Its form automatically ensures that
\equ{
\label{eq:IPF:ath_marginal}
\sum_{\ga=1}^\dimstatespace C_{a\ga}\prob p^{(nD+a)}_\config = \widehat m_a
}
right after fitting onto the $a$-th marginal, so that each term vanishes identically in the latter sum of \eqref{eq:IPF:VanishingRelation}, since $\prob p\in[\prob f]$ by assumption. Vanishing relation \eqref{eq:IPF:VanishingRelation} can be equivalently rewritten as
\equ{
\infdiv{\prob p}{\prob p^{(nD+a-1)}} = \infdiv{\prob p}{\prob p^{(nD+a)}} + \infdiv{\prob p^{(nD+a)}}{\prob p^{(nD+a-1)}}
\quads{\text{for}} \prob p\in\mathcal [\prob f]
~.
}
in terms of \textsc{kl} divergences of \ipf estimates. By induction we obtain after $n$ cycles 
\equ{
\infdiv{\prob p}{\prob p^{(0)}} - \infdiv{\prob p}{\prob p^{(nD)}} = 
\sum_{n'=0}^{n-1}\sum_{a=1}^D\infdiv{\prob p^{(n'D+a)}}{\prob p^{(n'D+a-1)}}
~.
}
Whenever starting from the uniform distribution $\prob p^{(0)}=\prob u$ and by definition of the equivalence class, where any structural zero appearing in the marginals and hence incorporated into $\prob p^{(nD)}$ would be also shared by all $\prob p\in[\prob f]$, the \textsc{kl} divergences on the l.h.s.\ are finite. Since the l.h.s.\ always remains bounded as $n \rightarrow\infty$, the sum of non-negative terms on the r.h.s.\ must be bounded, as well. Hence, we know by Cauchy criterion that there must exist $n^*\in\mathbb N$ so that 
\equ{
\infdiv{\prob p^{(nD+a)}}{\prob p^{(nD+a-1)}} < \epsilon 
\quads{\text{for}} n\geq n^* \qand a=1,...,D
~,
\nonumber
}
given any $\epsilon>0$.
In turn, this implies that \ipf estimates $\prob p^{(nD+a)}$ induce a Cauchy sequence in $[\prob f]$, thus establishing the existence of a generically real-valued limiting distribution  $\maxentP'$ in the sequence of \ipf estimates. Since each $\prob p^{(nD+a)}$   satisfies  by merit of  \eqref{eq:IPF:ath_marginal} the $a$-th marginal sum, cycling through all marginal constraints $a=1,...,D$ 
makes the limiting distribution $\maxentP'$ satisfy them all. Consequently, the limit $\maxentP'$ has to belong to $[\prob f]$. 
In particular, we conclude  after  finitely many steps that
\equ{
\hspace{1.3cm}\prob p^{(nD+a-1)} \approx \prob p^{(nD+a)}
\quads{\text{for}} n\geq n^* \qand a=1,...,D
\nonumber
}
within the desired accuracy $\epsilon$ (set e.g.\ by machine precision), which is obviously of practical importance. 

Eventually, it remains to verify that $\maxentP'$ is the \maxent distribution 
in $[\prob f]$. 
Given two distributions $\prob p,\prob p'\in[\prob f]$ it can be inductively shown that
\equ{
\label{eq:IPF:ith_Pythagorian2}
\sum_{\config=1}^\dimstatespace \left[\prob p_\config - \prob p'_\config\right] \log \prob p^{(nD+a)}_\config = 0
\quads{\text{for}} n=0,1,2,... \qand a=1,...,D
~.
}
Indeed, invoking \ipf update rule \eqref{eq:IPF_update} leads to 
\equ{
\sum_{\config=1}^\dimstatespace \left[\prob p_\config - \prob p'_\config\right] \log \prob p^{(nD+a)}_\config
=
\sum_{\config=1}^\dimstatespace  \left[\prob p_\config - \prob p'_\config\right] \log \prob p^{(nD+a-1)}_\config
+
\left(\log\widehat m_a-\log m^{(nD+a-1)}_a\right)\sum_{\config=1}^\dimstatespace  C_{a\ga}\left[\prob p_\config - \prob p'_\config\right]
= 0
~.
}
The second sum vanishes identically since  both $\prob p$ and $\prob p'$ sum to the observed $a$-th marginal from $\prob f$.
Starting from $n=0$ and $a=1$ the vanishing of the sum in the first term is trivial for uniform ansatz $\prob p^{(0)}_\config=\prob u_\ga=\dimstatespace^{-1}$, while it remains zero at the $(nD+a)$-th step by the inductive assumption, thus verifying the induction.
Taking $n \rightarrow \infty$ in  \eqref{eq:IPF:ith_Pythagorian2} and setting $\prob p'=\maxentP'$ (recall that the limiting distribution of \ipf belongs to $[\prob f]$) results into
\equ{
\sum_{\config=1}^\dimstatespace \left[\prob p_\config - \maxentP'_\config\right] \log \maxentP'_\config = 0
~.
}
Since this applies for arbitrary $\prob p\in [\prob f]$,  we recognize  from base Lemma~\ref{lemma:PythagoreanIdentity} that the limiting distribution $\maxentP'$ actually is of maximum entropy in $[\prob f]$ namely $\maxentP'=\maxentP$. This shows that \ipf always converges to the \maxent distribution concluding the proof of both Theorems~\ref{theorem:MaxEnt} and~\ref{theorem:IPF}.

\section{Variational formulation}
\label{ap:VarForm}

To relate to the widely used variational formulation we derive an explicit  form of \maxent distribution invoking the method  of Lagrange multipliers.
As variational approaches would fail to capture the optimal solution, if the extremum lied within a cusp of the Lagrangian, we are going to miss any microstates of zero probability. 
In the subspace $\mathcal A'$ where all microstates associated to structural zeros have been removed, we define the Lagrangian functional
\equ{
\mathcal L_\theta[\prob p] = H[\prob p] + \sum_{\ga=1}^{\vert\mathcal A'\vert} \prob p_\ga  - \sum_{a=1}^{D'}\theta_a\left(\sum_{\ga=1}^{\vert\mathcal A'\vert}  C'_{a\ga}\prob p_\ga -\prob m_a \right)
~,
}
parametrized by Lagrange multipliers $\theta_a\in\mathbb R$ to implement the provided phenomenological constraints 
\equ{
\label{eq:ReducedLinearSystem}
\sum_{\ga=1}^{\vert\mathcal A'\vert} C'_{a\ga}\prob p_\ga \overset{!}{=}\widehat m_a
\quads{\text{for}}  a=1,\ldots,D'~.
}
Here, $\mathbf C'$ denotes the reduced coefficient matrix and any constraints where $\widehat m_a=0$ have been correspondingly removed from $\widehat m$.
Extremizing first $\mathcal L$ w.r.t.\ $\prob p_\ga$ we obtain the familiar form of Boltzmann distribution,
\equ{
\maxentP_\ga = \exp \sum_{a=1}^{D'} \theta_a C'_{a\ga}
~.
}
Setting next the variation of $\mathcal L$ w.r.t.\ Lagrange multipliers to zero fixes the $\theta_a$'s via non-linear system
\equ{
\sum_{\ga=1}^{\vert\mathcal A'\vert} C'_{a\ga} \exp \sum_{b=1}^{D'} \theta_b C'_{b\ga} \overset{!}{=}\widehat m_a
\quads{\text{for}} a=1,\ldots, D'
\nonumber
} 
to obey the marginal constraints $m=\widehat m$. In total, we obtain after incorporating any structural zeros (due to $0\log0=0$ this does not change the extremum) the form 
\equ{
\label{eq:MaxEnt_PD}
\maxentP_\config = \left\lbrace
\begin{array}{ll}
\exp\sum\limits_{a=1}^{D'} \theta_{a} C'_{a\ga}~, & \config=1,...,\vert\mathcal A'\vert
\\[1ex]
0~, & \config=\vert\mathcal A'\vert+1,...,\dimstatespace 
\end{array}\right.
}
for the distribution with the  maximal differential entropy $H$ provided a consisted set of  marginal constraints.

Generically, the exponential part in the parametric form of  $\maxentP$ cannot be unambiguously fixed,  as long as the linear program~\ref{problem:MaxEnt_LP} has redundancies to begin with. Any choice of marginal constraints $\widehat m$ from empirical distribution $\prob f$ that spans the same equivalence class $[\prob f]$ (with or without structural zeros) would in principle lead to a different real-valued solution vector $\theta$, but of course to the same \maxent distribution $\maxentP$, as demonstrated in the previous section.  
In fact, the converse also holds: Any probability distribution $\hat{\prob q}\in[\prob f]$  of the form \eqref{eq:MaxEnt_PD} is the \maxent distribution. 
Taking an arbitrary $\prob p\in[\prob f]$ we have
\equ{
\sum_{\config=1}^\dimstatespace \left(\prob p_\config - \hat{\prob q}_\ga\right)\log\hat{\prob q}_\ga
= 
\sum_{\ga=1}^{\vert\mathcal A'\vert} \left(\prob p_\config - \hat{\prob q}_\ga\right)
\sum_{a=1}^{D'} \theta_a C'_{a\ga}
=
\sum_{a=1}^{D'} \theta_a \sum_{\ga=1}^{\vert\mathcal A'\vert} \left(C'_{a\ga}\prob p_\ga - C'_{a\ga}\hat{\prob q}_\ga\right)
 = 0
 ~,
 \nonumber
}
the inner sum in the l.h.s.\ of last equality vanishing separately for all $a$'s as both $\prob p$ and $\hat{\prob q}$ solve by assumption linear system  \eqref{eq:ReducedLinearSystem}. Applying Lemma~\ref{lemma:PythagoreanIdentity} we then conclude that $\hat{\prob q}$ has maximum entropy in $[\prob f]$, hence $\hat{\prob q}=\maxentP$.
In literature, such reparametrization invariance of the optimal solution and hence of meaningful quantities such as mutual information, odds and risk metrics is called a \textit{gauge symmetry}.

\bibliographystyle{paper}
{\small
\providecommand{\href}[2]{#2}\begingroup\raggedright\endgroup
}

\end{document}